\documentclass[]{elsarticle}
\usepackage{graphicx,url,hyperref,microtype,subcaption,graphbox,amsmath}
\usepackage{latexsym,courier,upgreek,xcolor,rotating,ulem}
\usepackage[onehalfspacing]{setspace}
\hypersetup{colorlinks,citecolor=blue,linkcolor=red,urlcolor=green}
\usepackage[displaymath,mathlines]{lineno}
\usepackage[margin=2cm]{geometry}
\modulolinenumbers[5]
\biboptions{numbers,sort&compress}

\begin{document}

\begin{frontmatter}
    \title{Approaching the continuum limit: effective rheology during a two-phase flow}

    \author[add1]{Subhadeep Roy}
    \ead{subhadeep.r@hyderabad.bits-pilani.ac.in}
    \author[add2]{Santanu Sinha}
    \ead{santanu.sinha@ntnu.no}
    \author[add3]{Alex Hansen}
    \ead{alex.hansen@ntnu.no}

    \address[add1]{Dept. of Physics, BITS Pilani Hyderabad Campus, Secunderabad, Telangana 500078, India}
    \address[add2]{PoreLab, Department of Physics, University of Oslo, N-0316 Oslo, Norway}
    \address[add3]{PoreLab, Department of Physics, Norwegian University of Science and Technology, N-7491 Trondheim, Norway}

    \begin{abstract}
        It is becoming increasingly clear that there is a regime in immiscible
        two-phase flow in porous media where the flow rate depends on the pressure drop
        as a power law with an exponent different than one. This occurs when the
        capillary forces and viscous forces both influence the flow.  At higher flow
        rates, where the viscous forces dominate, the flow rate depends linearly on the
        pressure drop.  The question we pose here is what happens to the non-linear
        regime when the system size is increased.  Based on analytical calculations
        using the capillary fiber bundle model and on numerical simulations using a
        dynamical network model, we find that the non-linear regime moves towards
        smaller and smaller pressure gradients as the system size grows.
    \end{abstract}
    
    \date{\today}
    
        
\end{frontmatter}


\makeatletter
\def\ps@pprintTitle{%
 \let\@oddhead\@empty
 \let\@evenhead\@empty
 \def\@oddfoot{}%
 \let\@evenfoot\@oddfoot}
\makeatother

\section{Introduction}
\label{intro}

In 1856, Darcy published  his famous treatise where the law that flow
rate is proportional to a pressure drop when a fluid flow through a
porous medium, was first presented \cite{d56}.  Eighty years later,
the Darcy law was generalized to the simultaneous flow of two
immiscible fluids by Wyckoff and Botset \cite{wb36}.  The basic idea
behind this generalization is that each fluid sees an available space
in which it can flow which consists of the pore space minus the space
the other fluid occupies.  Each fluid then obeys the Darcy law within
this diminished pore space.  This idea is clearly oversimplified.  It
remains to date, however with some important addenda such as the
incorporation of capillary effects \cite{l40}, the dominating tool for
simulations of immiscible two-phase flow in porous media. This is in
spite of numerous attempts over the years at improving this approach
or substitute it for an entirely new approach
\cite{hg90,hg93,hg93b,nbh11,gm14,kbhhg19,kbhhg19b,hb00,h06a,h06b,h06c,hd10,dhh12,vcp98,v12,v18,hsbkgv18,rsh19,rpsh22,hfss23,ph23,fsh23a}.

A simpler question may be posed when generalizing the Darcy equation to
immiscible two-phase flow in porous media.  Rather than asking for the flow rate
of each of the two fluids, how does the {\it combined flow\/} react to a given
pressure drop?  It has since Tallakstad et al.\ \cite{tkrlmtf09,tlkrfm09} did
their experimental study of immiscible two-phase flow under steady-state
conditions in a Hele-Shaw cell filled with fixed glass beads becomes
increasingly clear that there is a flow regime in which the flow rate is
proportional to the pressure drop with power different than one
\cite{gh11,rcs11,sh12,sh17,ydkst19,glbb19,fsrh21,fsh23}.  
That is, the two immiscible fluids flowing at the pore scale act at the
continuum scale as a single non-Newtonian fluid, or more precisely a
Herschel-Bulkley fluid where the effective viscosity depends on the shear rate,
and hence the flow rate, as a power law \cite{hb26}.

In the experimental setups that have been used, the flow rate of each fluid into
the porous medium is controlled and the pressure drop across the porous medium
is measured.  This leads to at least one of the fluids percolating even at very
low flow rates. At these low flow rates, the capillary forces are too strong for
the viscous forces to move the fluid interfaces, resulting in the standard
linear Darcy law prevailing.  As the flow rates are increased, Gao et al.\
\cite{glbb19} report a regime occurring where there are strong pressure
fluctuations but still the linear Darcy law is seen. Then, at even higher flow
rates, non-linearity sets in, and a power law relation between flow rate and
pressure drop is measured. This non-linearity may be associated with the gradual
increase in mobilized interfaces as the flow rates increase
\cite{tlkrfm09,sh12}. Lastly, at very high flow rates, the capillary forces
become negligible compared to the viscous forces, and again, the system reverts
to obey a linear Darcy law \cite{sh17}.

A simplified problem compared to that of immiscible two-phase flow in porous
media is that of bubbles flowing in a single tube
\cite{shbk11,xw14,lrth22,cfhws23}. Sinha et al.\ \cite{shbk11} studied a bubble
train in a tube with a variable radius assuming no fluid films forming. The main
result was that the time-averaged flow rate depends on the square root of the
{\it excess pressure drop,\/} that is the pressure drop along the tube minus a
depinning --- or threshold pressure $P_t$.  Xu and Wang \cite{xw14} also
identified a threshold pressure in their numerical simulations.  However, this
threshold pressure has a different character from that in the previous study: It
is the pressure drop at which contact lines start getting mobilized.  The
movement of the contact lines consumes energy leading to the effective
permeability dropping.  Xu and Wang \cite{xw14} suggest that this is the main
mechanism responsible for the non-linearity in the flow-pressure relationship.
Lanza et al.\ considered an immisible mixture of a non-Newtonian and a Newtonian
fluid moving along the tube \cite{lrth22}, whereas Cheon et al.\ considered a
mixture of compressible and incompressible fluids moving along the tube
\cite{cfhws23}. In both cases, a non-trivial power law dependence between the
flow rate and pressure drop. 

The question of whether there should be a threshold pressure or not in the
non-linear regime is an important one as assuming there to be one may alter
significantly the measured value of the exponent $\beta$ seen in the non-linear
regime where
\begin{equation}
\label{eq_nonlinear_darcy_1st}
Q\sim \begin{cases}
0\;, & {\rm if}\ |\Delta P|\le P_t\;, \\
(|\Delta P|-P_t)^\beta\;, & {\rm if}\ |\Delta P|> P_t \;,\\
\end{cases}
\end{equation}
where $Q=Q_w+Q_n$ is the volumetric flow rate consisting of the sum of
volumetric flow rates of the wetting fluid $Q_w$, and the non-wetting
fluid $Q_n$. $\Delta P$ is the pressure drop across the sample.
The value of $\beta$ varies in the literature.  Tallakstad et
al.\ \cite{tkrlmtf09,tlkrfm09} reported $\beta=1/0.54=1.85$ (in these
papers the inverse exponent was reported), Rassi et al.\ \cite{rcs11}
reported a range of values, $\beta=1/0.3=3.3$ to $\beta=1/0.45=2.2$,
and \cite{glbb19} reported $\beta=1/0.6=1.67$.  These results are
based on experiments and they all assume $P_t=0$.  Sinha et
al.\ \cite{sh17} report for their experiments $\beta=1/0.46=2.2$,
based on there is a threshold. Sinha and Hansen \cite{sh12} in
numerical work also assumes a threshold pressure based on a dynamic
network simulator \cite{jh12}, where fluid interfaces are moved
according to the forces they experience
\cite{amhb98,gvkh19,sgvh19,zmpcvz19}, and found $\beta=1/0.51=2.0$. The network representing the porous medium was here a disordered square lattice.  They followed this up with an effective medium calculation yielding $\beta=2$. 
Sinha et al.\ \cite{sh17} reported
$\beta=1/0.50=2.0$ to $\beta=1/0.54=1.85$ based on numerical studies
with reconstructed porous media using the same numerical model as in
\cite{sh12}. Yiotis et al.\ \cite{ydkst19} propose $\beta=3/2$ based
on numerical work and assuming the existence of a threshold pressure.
Recently Fyhn et al.\ \cite{fsh23} have studied a network model for a mixture of grains with opposite wetting properties with respect to the two immiscible fluids.  Depending on the filling ratio between the two grain types, there is a regime where there is no threshold pressure. They find an exponent $\beta=2.56$ in this regime.   

There is a lesson to be learned from the study of a very different
problem.  In 1993 M{\aa}l{\o}y et al.\ \cite{mwhr93} published an
experimental study where a rough hard surface was pressed into a soft
material with a flat surface, measuring the force as a function of the
deformation.  At first contact, the Hertz contact law was seen,
i.e.,\ the force depended on the deformation to the 3/2 power.  As the
deformation proceeded, a different power law emerged, however not in
the deformation but in the deformation minus a threshold deformation.
And here is the lesson: the threshold deformation was {\it not\/} the
deformation at first contact where the Hertz contact law was
seen. Transferring this result to the non-linear Darcy case, our point
is that the threshold pressure that shows up in the power law does
{\it not\/} have to be the pressure needed to get the fluids flowing.
The power law (\ref{eq_nonlinear_darcy_1st}) may be followed down to a
certain pressure difference larger than $P_t$. At this pressure
difference, there may then be a crossover to a different regime
controlled by different physics, e.g., a linear one as Guo et
al.\ \cite{glbb19} reported.

In this paper, we will discuss another aspect of the non-linear flow
regime which so far has not been touched upon.  So far, the system
sizes that have been used in establishing the existence of the
non-linear regime, even if the details are not yet sorted out, are
limited.  This applies both to the experimental and numerical
studies that have been published.  The question we pose here is: what
happens to the non-linear regime when the scale up the system, i.e.,
we go to the continuum limit? Does the threshold pressure $P_t$ remain
constant, increase or does it shrink away?  Does crossover to the
linear Darcy regime remain fixed at a given pressure gradient or changes?

Our conclusion, based on numerical evidence from the dynamic network
model \cite{amhb98,gvkh19,sgvh19} and on analytic calculation using
the capillary fiber bundle model \cite{rhs19}, is that the non-linear
regime shrinks away with increasing system size.

In the next section, we present a scaling analysis of the Darcy law
and the non-linear regime that sets the stage for the study that
follows.  We then turn in Section III to the capillary fiber bundle
model.  Section IV contains our numerical study based on scaling up
the square lattice. The last section contains a discussion of the
arguments presented earlier in the paper together with our conclusion.


\section{Scaling analysis}
\label{scaling}


We assume a porous medium sample that has length $L$ and an
transversal area $A$.  There is a pressure drop $\Delta P$ across it
and this generates a volumetric flow rate of $Q$.  When the flow rate is high so that capillary forces may be neglected, the constitutive
relation between $Q$ and $\Delta P$ is given by the Darcy law,
\begin{equation}
\label{eq_darcy}
Q=-M_d \Delta P\;,
\end{equation}
where $M_d$ is the mobility.  We introduce the Darcy velocity
\begin{equation}
\label{eq_darcy_vel}
v=\frac{Q}{A}\;,
\end{equation}
and the pressure gradient
\begin{equation}
\label{eq_press_grad}
p=\frac{\Delta P}{L}\;.
\end{equation}
The Darcy equation then takes the form
\begin{equation}
\label{eq_darcy_loc}
v=-m_d\ p\;,
\end{equation}
where
\begin{equation}
\label{eq_darcy_mob_loc}
m_d=\frac{M_d L}{A}\;.
\end{equation}
Equations (\ref{eq_darcy_loc}) and (\ref{eq_darcy_mob_loc}) are both
independent of the transversal area $A$ and the length $L$ of the
sample.

As has been described in the Introduction, there is a regime in which
the volumetric flow rate $Q$ depends on the pressure drop $\Delta P$
as a power law,
\begin{equation}
\label{eq_nonlin}
Q =-M_\beta\ {\rm sign}(\Delta P) \Theta(|\Delta P|-P_t)(|\Delta P|-P_t)^{\beta}\;,
\end{equation}
where $M_\beta$ is the non-linear mobility and $P_t$ is a threshold
pressure. Here $\Theta(|\Delta P|-P_t)$ is the Heaviside function
which is one for positive arguments and zero for negative arguments.
We use the Heaviside function to mark the end of the non-linear
regime when the pressure drop is lowered. 
There may be a crossover to a different regime before reaching
this lower cutoff \cite{glbb19}.

We have in the Introduction pointed out that the non-linear regime,
(\ref{eq_nonlin}), crosses over to the ordinary linear Darcy law
behavior above a maximum pressure difference, which we will call
$P_M$. In the following, we will assume that $P_t$ and $P_M$ have the
same dependence on the system sizes $A$ and $L$. We will support this
assumption in the next section where we study the capillary fiber
bundle model.

We express the non-linear Darcy law (\ref{eq_nonlin}) in terms of the
Darcy velocity and the pressure gradient,
\begin{equation}
\label{eq_nonlin_local}
v=-m_\beta\ {\rm sign}(p) \Theta\left(|p|-p_t\right)\left(|p|-p_t\right)^{\beta}\;,
\end{equation}
where
\begin{equation}
p_t=\frac{P_t}{L}\;,
\end{equation}
and
\begin{equation}
p_M=\frac{P_M}{L}\;.
\end{equation}
We then have that
\begin{equation}
\label{eq_nonlin_mob_loc}
m_\beta=\frac{M_\beta L^\beta}{A}\;.
\end{equation}

The continuum limit is reached by setting $A\sim L^{d-1}\to\infty$,
where $d$ is the dimensionality of the sample, and letting
$L\to\infty$.  In the Darcy regime, equations (\ref{eq_darcy}) to
(\ref{eq_darcy_mob_loc}), $v$, $p$ and the mobility $m_d$ are
independent of $L$.  The non-linear regime is different.  The
non-linear regime where the constitutive equation
(\ref{eq_nonlin_local}) applies, $v$ and $p$ are also independent of
$L$.  However, this is not the case for the threshold pressure $p_t$,
the crossover pressure $p_M$, and the mobility $m_\beta$.

\begin{figure}[ht]
    \centerline{\includegraphics[width=0.6\textwidth,clip]{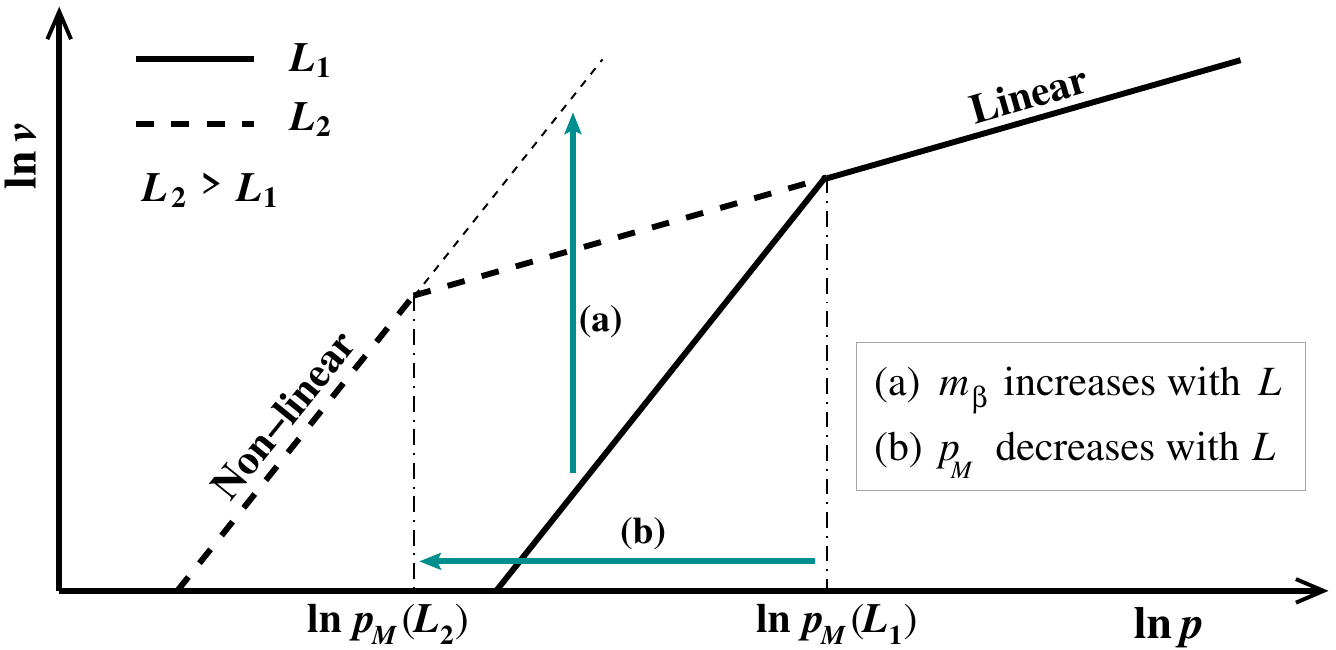}}
    
    \caption{We show $\ln v$ vs.\ $\ln p$ in both the linear range, equation
    (\ref{eq_darcy_loc}), and the non-linear range, equation
    (\ref{eq_nonlin_local}). Extrapolating the linear part of the curve to $\ln
    p=0$, it will cross the $\ln v$ axis at $\ln m_d$, where $m_d$ is the Darcy
    mobility (\ref{eq_darcy_mob_loc}). Extrapolating the non-linear part of the
    curve to $\ln p=0$, it will cross the $\ln v$ axis at $\ln m_\beta$, where
    $m_\beta$ is the non-linear mobility (\ref{eq_nonlin_mob_loc}). The linear
    mobility $m_d$ does not depend on the system size $L$. However, as we shall
    see, the non-linear mobility $m_\beta$ {\it grows\/} with increasing $L$,
    see arrow marked (a). This means that the crossover pressure $p_M$, where
    the linear and non-linear part of the curve $\ln v$ vs.\ $\ln p$ cross moves
    to the left in the figure, illustrated with arrow (b). Hence, $p_M$
    decreases with increasing $L$. We have set the threshold pressure $p_t$ to
    zero in this figure.}
    \label{fig0}
\end{figure}

We note that if $m_\beta\to\infty$ and $p_t\to 0$ as $L\to\infty$ as $A\sim
L^{d-1}\to\infty$, the non-linear regime {\it vanishes\/} in the continuum
limit. One may see this by sketching the Darcy law (\ref{eq_darcy_loc}) as a
straight line in a log-log plot of $v$ vs.\ $p$ as illustrated in figure
\ref{fig0}. The non-linear regime will give another straight line in this
diagram with slope $\beta$ when we ignore the threshold correction $|p|-p_t\to
|p|$.  We have $\beta>1$ so that the two lines cross each other with the
non-linear line below the Darcy line to the left and above to the right.  The
system follows the lowest of the two lines for any $|p|$.  If now the non-linear
$m_\beta$ mobility increases with increasing $L$, the cross point between the
two lines moves to the left, with the result that the non-linear regime moves to
lower and lower value of the pressure gradient $p$ as seen in figure \ref{fig0}.  

The reader should note a subtlety here. If $m_\beta\to\infty$ as $L\to\infty$
and $A\to\infty$, then we must have the crossover pressure $p_M\to 0$ as a
consequence.  This makes it unnecessary to measure $p_M$ --- a quantity that is
very difficult to measure with any accuracy; it is enough to measure $m_\beta$,
and not $p_M$.


\section{Capillary Fiber Bundle Model}
\label{fiber}


We now consider the capillary fiber bundle model \cite{s53,s74} as this is a
system that can be solved analytically. This model consists of $N$ parallel
capillary tubes of equal length $L$. The average transversal area of each tube
is $a$ so that $A=Na$. The radius of each tube varies with the position along
its axis.  We follow the approach of Sinha et al.\ \cite{shbk11} assuming that
the radius $r$ varies as
\begin{equation}
\label{eq_radius}
r(x)=\frac{r_0}{1-b\cos(2\pi x/l)}\;,
\end{equation}
where $r_0=\sqrt{a/\pi}$ is the average radius, $0<x<L$ is the position along
the capillary fiber and $l$ is the period of the radius variation. The capillary
tube is filled with bubbles. Neither of the two immiscible fluids wet the tube
walls completely so that there are no films.  We now focus on one bubble of the
less-wetting fluid.  The bubble is limited by interfaces at $x_I < x_F$ so that
the length of the bubble is $\Delta x_B=x_F-x_I$ and the position of its center
of mass is $x_B=(x_I+x_F)/2$.  The capillary pressure drop at $x=x_I$ is
\begin{equation}
\label{eq_capillary_xi}
\frac{2\sigma}{r(x_I)}=+\frac{2\sigma}{r_0}\ \left[1-b\cos\left(\frac{2\pi}{l}x_I\right)\right]\;,
\end{equation}
and the capillary pressure drop at $x_f$ is
\begin{equation}
\label{eq_capillary_xf}
\frac{2\sigma}{r(x_F)}=-\frac{2\sigma}{r_0}\ \left[1-b\cos\left(\frac{2\pi}{l}x_F\right)\right]\;,
\end{equation}
where $\sigma$ is the surface tension. The sum of these two forces gives the
capillary force on the bubble,
\begin{equation}
\label{eq_capillary_xb}
p_c(x_B)=-\frac{4b\sigma}{r_0}\ \sin\left(\frac{\pi}{l}\Delta x_B\right)\sin\left(\frac{2\pi}{l}x_B\right)\;.
\end{equation}
Suppose now there are $k$ bubbles per unit length in the capillary tube so that
it contains $K=kL$ bubbles. At the time $t$ their centers of mass are positioned
at $x_i(t)$, where $1\le i\le K$.  The equation of motion for bubble number $i$
is
\begin{equation}
\label{eq_motion_bubble_i}
\dot{x}_i=-\frac{r_0^2}{8L\mu_{\rm eff}}\left[\Delta P+\sum_{i=1}^K \frac{4b\sigma}{r_0}
\sin\left(\frac{\pi}{l}\Delta x_i\right)\sin\left(\frac{2\pi}{l}x_i\right)\right]\;,
\end{equation}
where $\mu_{\rm eff}=\mu_n\sum_{i=1}^K \Delta x_i+\mu_w (L-\sum_{i=1}^K \Delta
x_i)$, in which $\mu_n$ is the viscosity of the non-wetting fluid and $\mu_w$ is
the viscosity of the wetting fluid. We now introduce relative coordinates
$\delta x_i=x_i-x_0$ where $x_0$ is some chosen point along the abscissa. We
have that $\dot{x}_0= \dot{x}_1=\cdots =\dot{x}_K$. This implies that $\delta
x_i=0$ for all $i$.  We may then write the $K$ equations of motion
(\ref{eq_motion_bubble_i}) as a single equation
\begin{equation}
\label{eq_motion_bubble_0}
\dot{x}_0=-\frac{r_0^2}{8L\mu_{\rm eff}}\left[\Delta P+\Gamma_s \sin\left(\frac{2\pi}{l}x_0\right)
+\Gamma_c \cos\left(\frac{2\pi}{l}x_0\right)\right]\;,
\end{equation}
where
\begin{equation}
\label{eq_gamma_s}
\Gamma_s=\frac{4b\sigma}{r_0}\ \sum_{i=1}^K \sin\left(\frac{\pi}{l}\Delta x_i\right)
\sin\left(\frac{\pi}{l}\delta x_i\right)\;,
\end{equation}
and
\begin{equation}
\label{eq_gamma_c}
\Gamma_c=\frac{4b\sigma}{r_0}\ \sum_{i=1}^K \sin\left(\frac{\pi}{l}\Delta x_i\right)
\cos\left(\frac{\pi}{l}\delta x_i\right)\;,
\end{equation}
Let us set
\begin{equation}
\label{eq_capillary_thresh}
P_t=\sqrt{\Gamma_s^2+\Gamma_c^2}\;,
\end{equation}
and introduce the non-dimensional variables for $x_0$ and $t$,
\begin{equation}
\label{eq_theta}
\theta=\frac{2\pi}{l}x_0\;,
\end{equation}
and
\begin{equation}
\label{eq_time}
\tau=\frac{\pi r_0^2P_t}{4Ll\mu_{\rm eff}}\ t\;.
\end{equation}
Hence, equation (\ref{eq_motion_bubble_0}) becomes
\begin{equation}
\label{motion_bubble_theta}
\dot{\theta}=\frac{\Delta P}{P_t}-\sin\left(\theta+\theta_t\right)\;,
\end{equation}
where
\begin{equation}
\label{eq_theta_t}
\tan(\theta_t)=\frac{\Gamma_s}{\Gamma_c}\;.
\end{equation}
We see from this equation that $|\Delta P|$ must be larger than $P_t$ for the
bubbles to move in the capillary tube; $P_t$ is a {\it threshold pressure.\/}

(In references \cite{shbk11} and \cite{rhs19} there is an error in identifying
the mathematical form of the threshold pressure.  This error has no impact on
the results there.)

We now assume we scale $L$ in such a way that $k=K/L$ remains constant.  How
will $P_t$ scale with $L$?  Since the number of interfaces increase linearly
with $L$, one may be tempted to believe that $P_t$ scales with $L$. However, the
interfaces come in pairs, one for each bubble, and the capillary pressure drops
across the interfaces come with opposite signs.  Hence, the capillary pressure
$p_c(x_B)$ in equation (\ref{eq_capillary_xb}) can have {\it either\/} sign
depending on the size and position of the bubble, $\Delta x_B$ and $x_B$. With
$K$ bubbles, $\Gamma_s$ and $\Gamma_c$ are sums of factors that have random
signs; we are dealing with random walks. As a consequence, we have that
\begin{equation}
\label{eq_capillary_scale_L}
P_t \sim \sqrt{L}\;.
\end{equation}
A more general version of this argument has been presented in \cite{ffh22}. 

We now bring together $N$ of these capillary fibers to form a bundle
\cite{rhs19}. The fibers have radii $r_0$ drawn from some probability
distribution. Since the thresholds $P_t$ are inversely proportional to $r_0$, we
will consider the corresponding threshold probability distribution. We follow
\cite{rhs19} and consider first the cumulative probability
\begin{equation}
\label{eq_cumulative_0}
\Pi(P'_t)=\left\{\begin{array}{ll}
                                0       & \mbox{, $P_t \le 0$\;,}\\
                                \frac{P'_t}{P_M} & \mbox{, $0 < P'_t \le P_M$\:,}\\
                                1       & \mbox{, $P_t > P_M$\;,}\\
              \end{array}
       \right.
\end{equation}
where $P_M$ is the maximum threshold.  Note the change in notation: The
threshold associated with a given capillary fiber is $P'_t$. We reserve $P_t$
for the threshold pressure the whole capillary fiber bundle. Averaging the
equation of motion (\ref{eq_motion_bubble_0}) for each fiber in the bundle then
gives \cite{rhs19}
\begin{equation}
\label{eq_capillary_0_motion}
Q=-\frac{aA}{32\mu_{\rm eff}L}\left|\frac{\Delta P}{P_M}\right|\Delta P
\end{equation}
when $|\Delta P| \le P_M$. Hence, the threshold pressure $P_t=0$ when the
threshold distribution for the individual fibers is given by
(\ref{eq_cumulative_0}). Hence, we have that 
\begin{equation}
\label{eq_Mbeta_fb}
M_\beta=\frac{aA}{32\mu_{\rm eff} P_M L}\;.
\end{equation}

In terms of the Darcy velocity $v$ and the pressure gradient $p$, this
expression becomes
\begin{equation}
\label{eq_capillary_0_darcy}
v=-\frac{a}{32\mu_{\rm eff}\ p_M} \ |p|\ p=-m_\beta \ |p|\ p\;,
\end{equation}
where $p_M=P_M/L$. Hence, $\beta=2$. We see that $m_\beta$ has the same 
form as in equation (\ref{eq_nonlin_mob_loc}),
\begin{equation}
\label{eq_mbeta_fbundle}
m_\beta=\frac{a}{32\mu_{\rm eff}\ p_M}=\frac{M_\beta L^2}{A}\frac{1}{p_M}\;.
\end{equation}

$P_M$ is the threshold pressure for getting the fluid in the most
difficult fiber to flow. Hence, we will have that
\begin{equation}
\label{eq_capillary_0_pm}
p_M=\frac{P_M}{L} \sim \frac{1}{\sqrt{L}}\;,
\end{equation}
from equation (\ref{eq_capillary_scale_L}), and as a consequence
\begin{equation}
\label{eq_capillary_0_mbeta}
m_\beta\sim L^{1/2}\;.
\end{equation}
It is important to note that $P_t=0$ in this fiber bundle.  Thus, we have
$p_t=0$ and $m_\beta\to\infty$ in the limit $A\to\infty$ and $L\to\infty$: The
non-linear behavior disappears in the continuum limit, see figure \ref{fig0}.

We now consider the cumulative threshold probability \cite{rhs19}
\begin{equation}
\label{eq_capillary_1_prob}
\Pi(P_t)=\left\{\begin{array}{ll}
                                0       & \mbox{, $P'_t \le P_t$\;,}\\
                                \frac{P'_t-P_t}{P_M-P_t} & \mbox{, $P_t < P'_t \le P_M$\:,}\\
                                1       & \mbox{, $P'_t > P_M$\;,}\\
              \end{array}
       \right.
\end{equation}
noting that such a distribution is more realistic than one where the minimum
threshold is zero, see the distribution in equation (\ref{eq_cumulative_0}).
This is so since a zero threshold would mean that there is a possibility for an
infinite radius $r_0$ in equations (\ref{eq_capillary_xi}) and
(\ref{eq_capillary_xf}).  

The flow rate is in this case given by
\begin{equation}
\label{eq_capillary_1_q}
Q=-\frac{aA\ {\rm sign}(\Delta P)}{3\sqrt{2}\pi\mu_{\rm av} L}\frac{\sqrt{P_t}}{(P_M-P_t)}(|\Delta P|-P_t)^{3/2}\;,
\end{equation}
for $|\Delta P|$ close to but larger than the threshold $P_t$. In terms of the
Darcy velocity and pressure gradient, this expression becomes
\begin{eqnarray}
\label{eq_capillary_1_v}
v&=&-\frac{a\ {\rm sign}(p)}{3\sqrt{2}\pi\mu_{\rm av}}\frac{\sqrt{p_t}}{(p_M-p_t)}(|p|-p_t)^{3/2}\nonumber\\
&=&- m_\beta\ {\rm sign}(p)\ (|p|-p_t)^{3/2}\;,
\end{eqnarray}
where we have defined
\begin{equation}
\label{eq_capillary_1_pm}
p_t=\frac{P_t}{L}\;.
\end{equation}
Since $P_t$ is the threshold pressure for the capillary fiber with the smallest
threshold in the bundle, we must have
\begin{equation}
\label{eq_capillary_1_pm_1}
p_t\sim \frac{1}{\sqrt{L}}\;,
\end{equation}
from equation (\ref{eq_capillary_scale_L}). Combined with
(\ref{eq_capillary_0_pm}), we find
\begin{equation}
\label{eq_capillary_1_mbeta}
m_\beta \sim L^{1/4}\;.
\end{equation}
Hence, we find that $p_t\to 0$ and $m_\beta\to\infty$ in the limit $A\to\infty$
and $L\to\infty$: The non-linear behavior disappears also in this case in the
continuum limit.

Even though, we have found that $m_\beta$ to increase with $L$ based on the
capillary fiber bundle model, we believe this result to be generally applicable.
The reason for this is that the fluctuations of surface tension of the
interfaces keeping the fluids in place scale more slowly than the pressure
gradient. This is a mechanism that will be present also in porous media, and not
just in the capillary fiber bundles.     


\section{Numerical results based on a dynamic network model}
\label{numerical}


We base our simulations on the dynamic network simulator described in
\cite{amhb98,gvkh19,sgvh19}. It consists of interfaces that span the pores and
move according to the pressure gradient they experience. Hence, no wetting films
occur in the simulations. We use a square lattice oriented at $45^\circ$ to the
average flow direction. We assume periodic boundary conditions both in the
direction orthogonal to the average flow direction and in the direction parallel
to the average flow.

\begin{figure}[t]
    \centerline{\hfill
    \includegraphics[width=0.32\textwidth,clip]{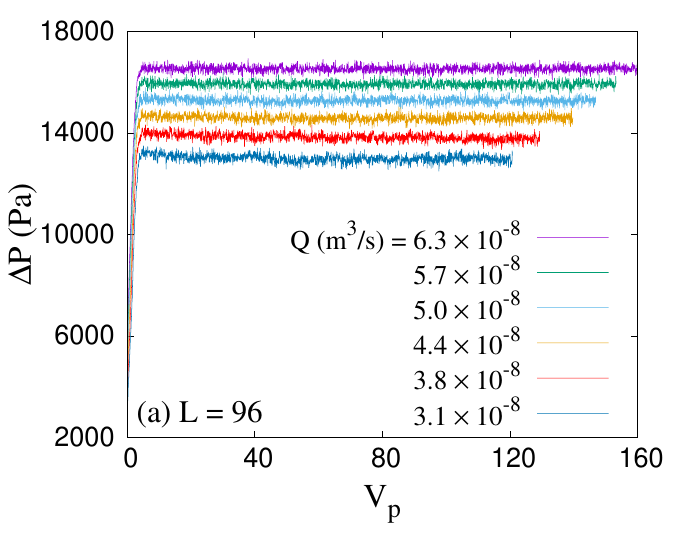}\hfill
    \includegraphics[width=0.32\textwidth,clip]{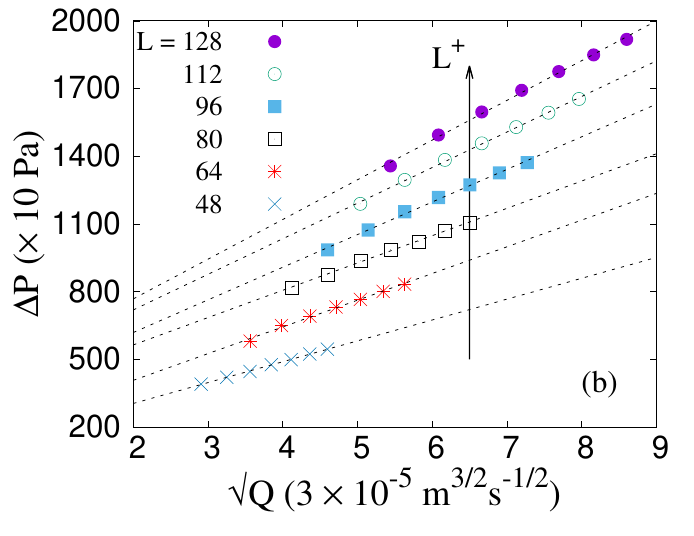}\hfill
    \includegraphics[width=0.32\textwidth,clip]{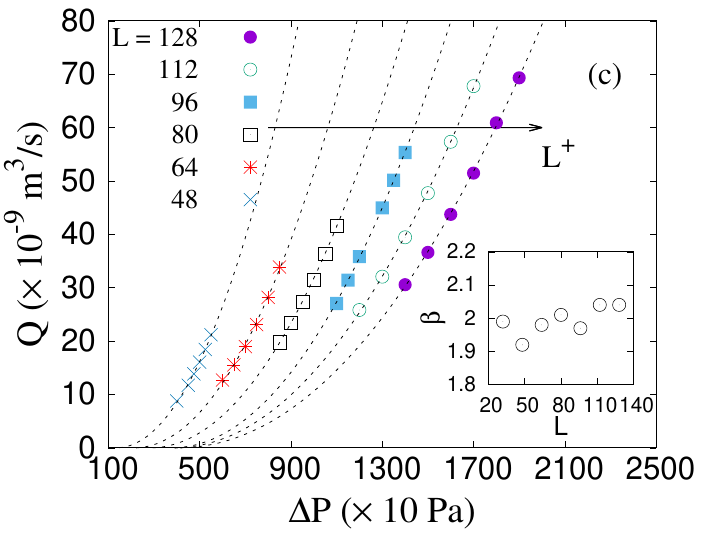}\hfill}
    \centerline{\hfill
    \includegraphics[width=0.32\textwidth,clip]{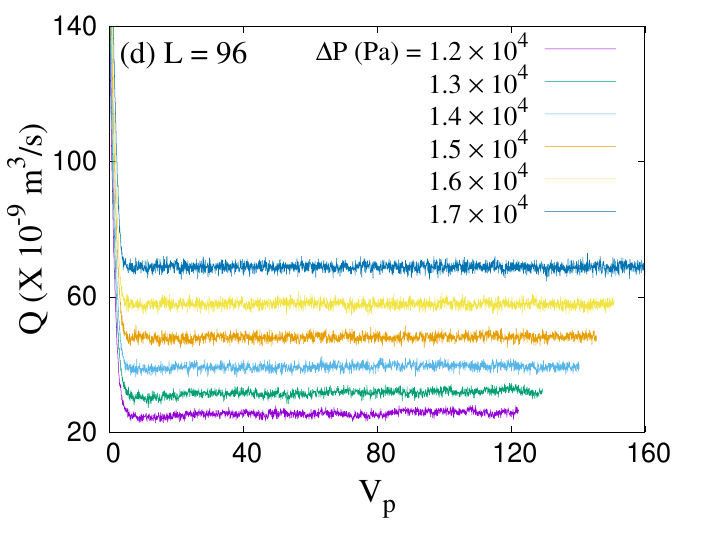}\hfill
    \includegraphics[width=0.32\textwidth,clip]{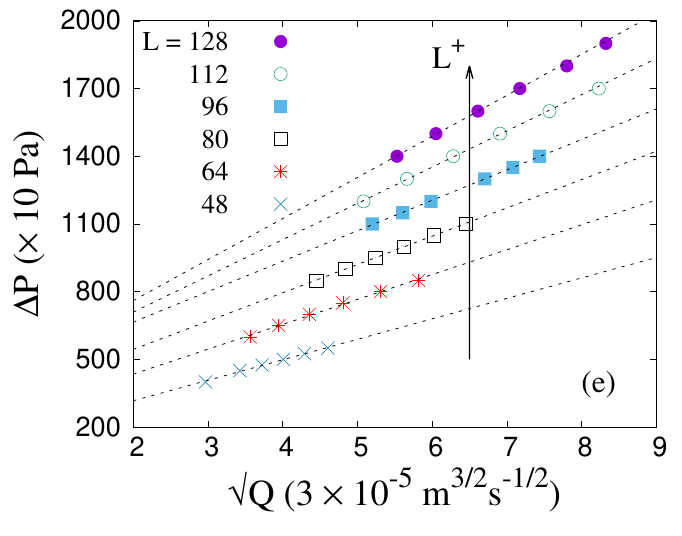}\hfill
    \includegraphics[width=0.32\textwidth,clip]{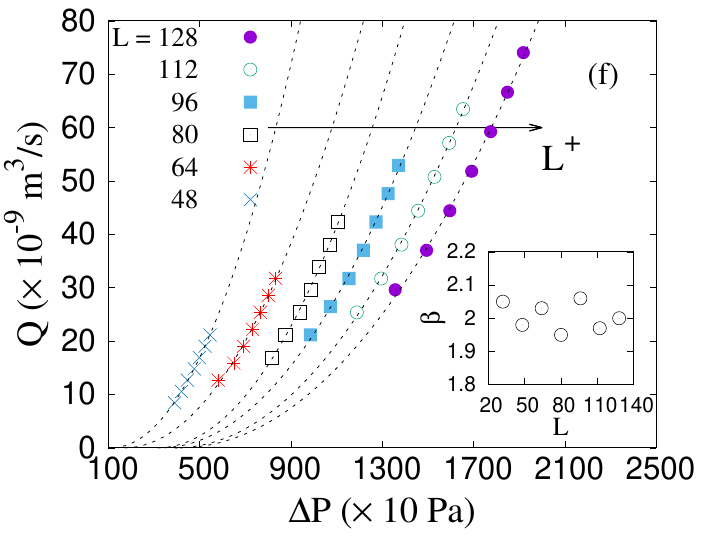}\hfill}
    \caption{The
    upper panel of the figure corresponds to the constant flow rate while the
    lower panel corresponds to constant pressure gradient. The size of the
    network used is $96 \times 96$. The saturation $S_w$ value is kept constant
    at 0.5. (a) \& (d) At a constant flow rate
    ($3.0\times10^{-8}<Q<6.5\times10^{-8}$ m$^3$/s) or pressure gradient
    ($1.2\times10^4<\Delta P<1.7\times10^4$ Pa), $\Delta P$ and $Q$ gradually
    approaches the steady-state value with increasing pore volumes $V_p$. (b) \&
    (e) We assume $\beta=2.0$. The figures show the variation of $\Delta P$ with
    $\sqrt{Q}$ at a constant flow rate (upper) and constant pressure gradient
    (lower). For both figures the system sizes from up to down are 128, 112, 96,
    80, 64, and 48. As the size of the system is increased both the slope of the
    straight line and the intercept on the ordinate increases. The value of
    $P_t$ and $M_{\beta}$ can be extracted from the intercept of the straight
    the line on the ordinate and its slope respectively (see equation
    \ref{eqn1.02}).  (c) \& (f) $\beta$ is treated to be a fitting parameter and
    the numerical results are fitted with the equation (\ref{eq_nonlin}) to find
    $\beta$, $M_{\beta}$ and $P_t$. The system sizes used here are the same as
    (b) and (e). The fitted $\beta$ value is observed to be close to $2.0$
    (shown in the inset).}
    \label{fig1}
\end{figure} 

The square lattices we have used range in size between $48 \times 48$ and $208
\times 208$. All the links are of length $l=10^{-3}$ m with its average radius
$r$ chosen randomly between $0.1l$ and $0.4l$. The simulation is carried out at
both constant flow rate $Q$ and constant pressure gradient $\Delta P$, kept at a
certain low value so that the capillary forces dominate and the relationship
between $Q$ and $\Delta P$ is non-linear.  For system sizes $L=48$, 64, 80, 96,
112, 128, 144, 160, 176, 192, and 208 we have used respectively 20, 20, 15, 15
10, 10, 8, 5, 3, 3, and 3 realizations. We set the surface tension $\sigma$ to
the value 0.03 or 0.01 N/m. While calculating the flow rate, instead of assuming
a cross-section, we summed up the flow rate for all links and divided it by the
total number of links.

\begin{figure}[ht]
    \centerline{\hfill
    \includegraphics[width=0.45\textwidth,clip]{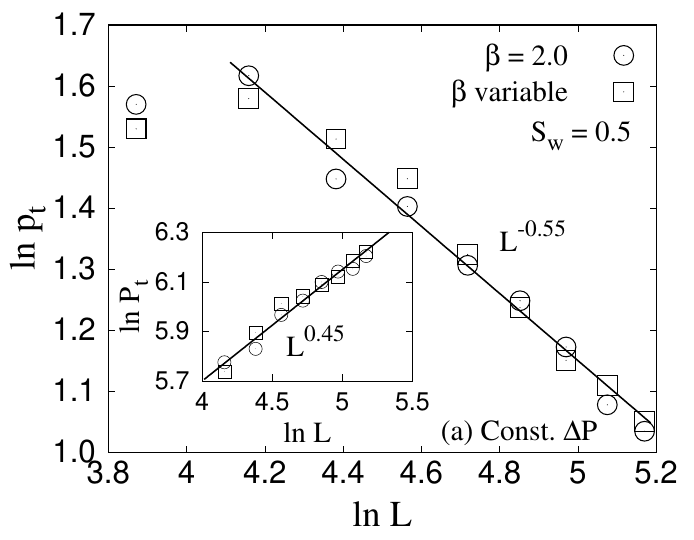}\hfill
    \includegraphics[width=0.45\textwidth,clip]{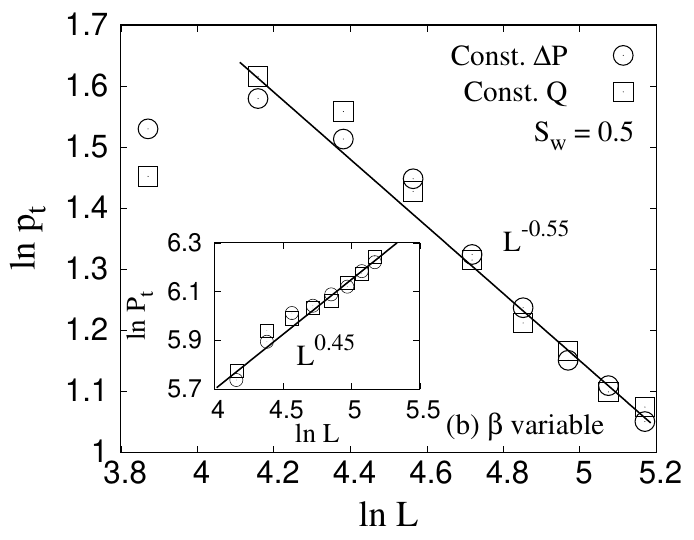}\hfill} 
    \caption{$p_t$
    as a function of $L$ where $L$ ranges from 48 to 176 is shown for (a)
    constant pressure gradient and $\beta=2.0$ as well as treating $\beta$ as a
    fitting parameter; (b) $\beta$ as a fitting parameter for both constant
    pressure and constant flow rate. The inset in both figures shows the size
    effect for $P_t$ under the same conditions. The saturation $S_w=0.5$ in all
    cases.}
    \label{fig2}
\end{figure} 
 
Figure \ref{fig1} shows the relation between the pressure gradient and the flow
rate when the model reaches the steady state. The upper panels of the figure
correspond to constant $Q$ while the lower panels show the results for constant
$\Delta P$.  We show in figure \ref{fig1}(a) pressure difference $\Delta P$ as a
function of injected pore volumes when keeping $Q$ constant and in figure
\ref{fig1}(d) $Q$ as a function of injected pore volumes when keeping $\Delta P$
constant.  We see that in both cases, within a few injected pore volumes the
system reaches a steady state.  All data are collected after the system reaches
a steady state. For the flow rates shown the system is well within the
non-linear region where the equation (\ref{eq_nonlin}) applies.

In order to calculate $P_t$ for a system size $L$ we have adopted two different
methods. For the first one we have assumed the mean-field solution from Sinha
and Hansen \cite{sh12}, setting $\beta=2$ in equation (\ref{eq_nonlin}). For the
second method, we keep $\beta$ free as a fitting parameter and the numerical
results are fitted with equation (\ref{eq_nonlin}) with variables $P_t$,
$M_{\beta}$ and $\beta$. We do not measure the crossover pressure $P_M$ where
the non-linear relation (\ref{eq_nonlin}) is replaced by the Darcy law
(\ref{eq_darcy}).  As we have already observed at the end of Section II, this is
not necessary when we determine $P_t$ and $M_\beta$.

\begin{figure}[]
  \centerline{\hfill
    \includegraphics[width=0.45\textwidth,clip]{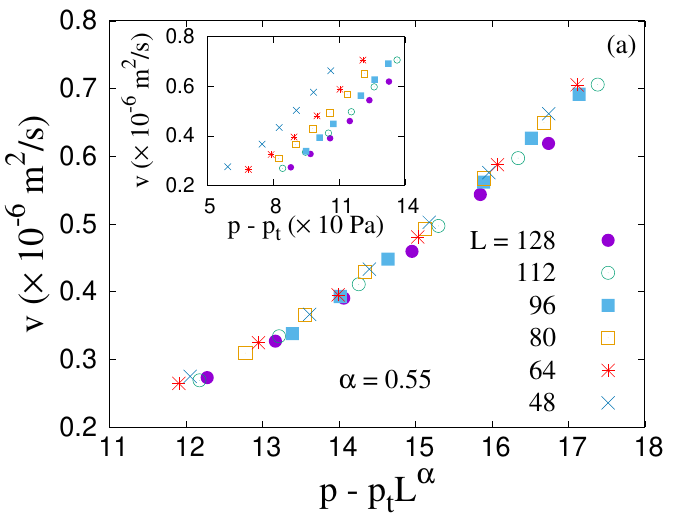}\hfill
    \includegraphics[width=0.45\textwidth,clip]{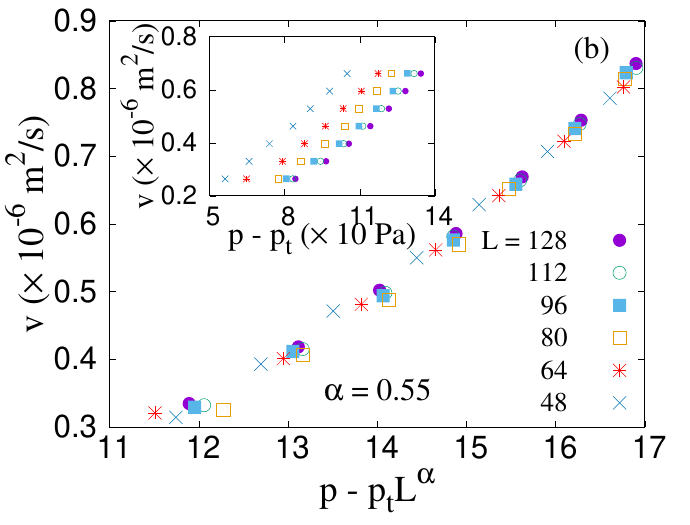}\hfill}
  \caption{Darcy velocity $v=Q/L$ plotted against
    $p-p_tL^\alpha=\Delta P/L-(P_t/L^{1+\alpha})L^\alpha$, where we
    have set $\alpha=0.55$, thus producing data collapse.  We assumed
    $\beta$ to be a fitting parameter. We furthermore set
    $\mu_n/\mu_w=1.0$ and $S_w=0.5$ respectively. The study was
    carried out for (a) constant pressure gradient and (b) constant
    flow rate.}
\label{fig3}
\end{figure} 

{\it Constant $\beta= 2$}: In the capillary force dominated region, if
we assume $\beta=2$, we get from equation (\ref{eq_nonlin}) that
\begin{equation}
\label{eqn1.02}
\Delta P \sim \sqrt{\frac{Q}{M_\beta}}+P_t\;,
\end{equation}
when taking into account the sign of $\Delta P$ used in the simulation.  Figures
\ref{fig1}(b) and (e) show how the pressure gradient $\Delta P$ behaves with
$\sqrt{Q}$ for constant flow rate and constant pressure gradient respectively.
In both cases, we observe a straight line whose intercept on ordinate gives the
value of $P_t$. As we increase $L$, the slope of the straight line as well as
the intercept $P_t$ increases. $M_{\beta}$ can be extracted from the slope of
this straight line.\\

{\it $\beta$ as fitting parameter}: Next, we have kept $\beta$ as a free
parameter and the numerical results are fitted with the equation
(\ref{eq_nonlin}). The fitted results are shown by dotted lines in figure
\ref{fig1}(c) and (f). The inset in the same figure shows the $\beta$ values for
different system sizes. The variation in $\beta$ values show that the mean-field
approximation is valid for our numerical results and $\beta$ has a value close
to 2.0.
 
\begin{figure}[t]
  \centerline{\hfill
  \includegraphics[width=0.45\textwidth,clip]{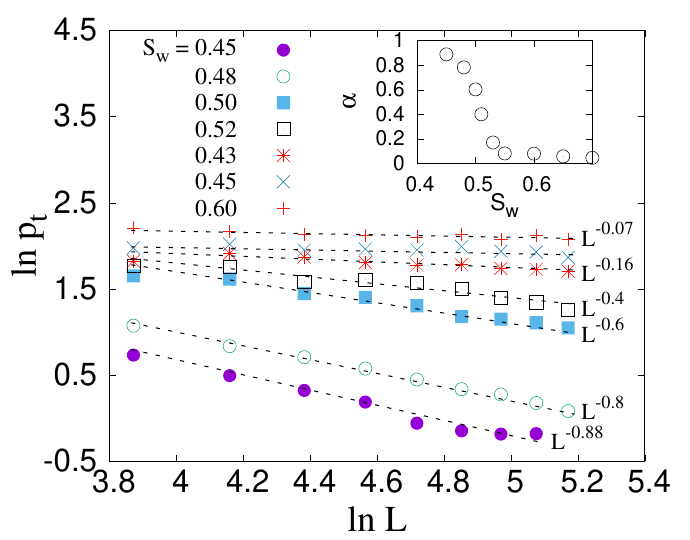}\hfill} 
\caption{Here we
  show $p_t=P_t/L$ as a function of $L$ for $L=48$ to $L=176$, $\mu_n/\mu_w=1.0$
  and for different values of $S_w$. The behavior is consistent with equation
  (\ref{eq_pt_num}). The exponent $\alpha$ is a strong function of $S_w$.
  However, all values of $\alpha$ are negative so that $p_t\to 0$ as
  $L\to\infty$.}
\label{fig4}
\end{figure} 

We now discuss the size effect of the threshold pressure $p_t=P_t/L$. In figure
\ref{fig2}(a) we show $p_t$ as a function of $L$ for constant pressure gradient
$\Delta P$ for the following two cases:  $\beta=2$, as well as when we keep
$\beta$ as an independent fitting parameter. In both cases, a scale-free decay
of $p_t$ is observed with $L$. Figure \ref{fig2}(b) shows the same power law
decay  for both constant $\Delta P$ and constant flow rate $Q$ with $\beta$
being treated as an independent fitting parameter. We find in all cases
\begin{equation}
\label{eq_pt_num}
p_t\sim L^{-\alpha}\;,
\end{equation}
where $\alpha=0.55$. We will, however, demonstrate later on that $\alpha$
depends on the saturation $S_w$.

Another way of displaying the dependence of the threshold pressure $p_t$ on the
system size $L$ is to plot the Darcy velocity $v$ as a function of
$p-p_tL^\alpha$. We should then observe data collapse for different values of
$L$. This is precisely what we observe in figure \ref{fig3}. We note that
whether we keep the pressure drop $\Delta P$ or the flow rate $Q$ constant, the
results are quite similar.  In light of this behavior, we will only consider the
constant pressure drop scenario in the following.  We will also in the following
keep $\beta$ as a free parameter.

The dependence of $p_t$ on saturation $L$ for various saturation $S_w$ is shown
in figure \ref{fig4}.  We observe $\alpha$ to remain constant at a low value for
$S_w>0.55$. In the region $0<S_w<0.55$, $\alpha$ increases quickly with
decreasing saturation. The variation $\alpha$ with $S_w$ is shown in the inset
of figure \ref{fig4}. In all cases, $\alpha$ is positive so that $p_t\to 0$ as
$L\to\infty$.

These results show that the capillary fiber bundle model which predicts
$\alpha=1/2$ does not capture the full mechanisms behind the scaling we observe.
We will return to this in the concluding section.

We now turn to the mobility $M_\beta$ and $m_\beta$ defined in equations
(\ref{eq_nonlin}) and (\ref{eq_nonlin_mob_loc}) respectively.  Figure \ref{fig5}
shows the size effect for both $M_{\beta}$ and $m_\beta$.
\begin{equation}
\label{eqn1.06}
M_{\beta} \propto L^{-\eta}
\end{equation}
where $\eta$ has values 0.78 ($S_w=0.53$), 0.82 ($S_w=0.50$) and 0.75
($S_w=0.48$), hence the dependence on saturation.  From equation
(\ref{eq_nonlin_mob_loc}), we have that
\begin{equation}
\label{eq_nonlin_tilde_mbeta-2}
m_\beta=\frac{M_\beta}{L} L^\beta\sim L^{\beta-1-\eta}\;,
\end{equation}
where we have used that $A=L$ for the two-dimensional networks we use.  
With the value $\beta=2.0$, we find that $\beta-1-\eta$ is larger than zero for
all observed $\eta$-values. More specifically, we find $\beta-1-\eta=0.22$,
$0.18$ and $0.25$ respectively.  We show these results in figure \ref{fig5}.  

We note how close the exponents measured in figure \ref{fig5}b are to the
capillary fiber bundle model, equation (\ref{eq_capillary_1_mbeta}), where an
exponent 1/4 was found. 


\section{Discussion and conclusion}
\label{conclusion}


We have in this paper posed the question: Does the non-linear regime where the
flow rate depends on the pressure drop through a power law with exponent
different expand its range of validity, diminish it or stay the same?  We have
used two approaches to answer this question. The first one is to solve the
capillary fiber bundle model. In doing so, we find that indeed the non-linear
regime shrinks away with increasing system size.  The reason for this is that
the crossover pressure that defines the border between the non-linear regime and
the linear Darcy regime moves toward zero with increasing system size. This, in
turn, is a result of this threshold pressure $p_M$ is a sum of factors that
appear with random signs, thus rendering it into a random walk process.  The
mobility $m_\beta$ depends on the inverse threshold pressure to a power. This
ensures that it {\it increases\/} when the threshold pressure decreases, a
necessary and sufficient condition for the non-linear regime to shrink away.

\begin{figure}[t]
  \centerline{\hfill
  \includegraphics[width=0.45\textwidth,clip]{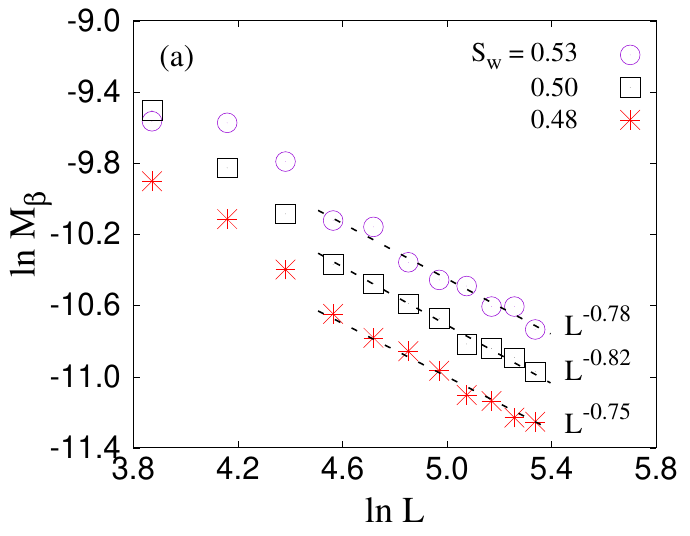}\hfill
    \includegraphics[width=0.45\textwidth,clip]{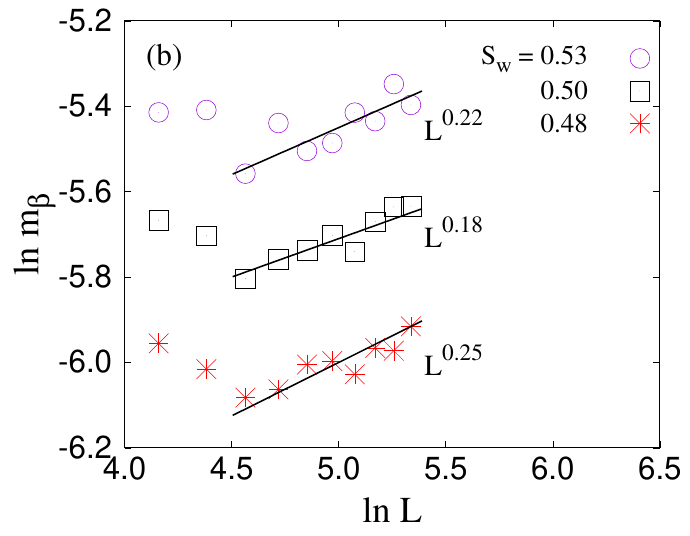}\hfill}
  \caption{The mobility $M_{\beta}$ defined in equation
    (\ref{eq_nonlin}) scales with system size $L$, ranging from $L=48$
    to $L=208$, as described in equation (\ref{eqn1.06}).  The scaled
    mobility (\ref{eq_nonlin_mob_loc}) then scales as
    $m_{\beta}=M_{\beta}L^{\beta-1} \sim L^{\beta-1-\eta}$.  Since
    $\eta<1$ and $\beta \approx 2.0$, $m_{\beta}$ increases with
    increasing $L$. We set $\mu_n/\mu_w=1.0$ here.}
\label{fig5}
\end{figure} 

We find the same qualitative behavior in the dynamic network model we
then employ: the threshold pressure $p_t$ shrinks and the mobility
$m_\beta$ increases with increasing system size.  Both quantities
depend on the system size according to a power law.  We find that the 
exponents depend weakly on the saturation $S_n$. However, they are quite close
to the values found in the capillary fiber bundle model when we assume
that the capillary threshold $P_t$ distribution does not go all the way to zero,
see equation (\ref{eq_cumulative_0}), a feature also present in the dynamic 
pore network model.  Compare the exponents observed in figure \ref{fig5} with the
scaling found for the mobility $m_\beta$ for the capillary fiber bundle model,
equation (\ref{eq_capillary_1_mbeta}).

We urge that experiments are done in order to move beyond the
theoretical and numerical considerations presented here with their
obvious limitations.

An understanding of the non-linear Darcy regime is very important as
it occurs right in the parameter range relevant for many industrial
situations such as oil recovery, water flow in aquifers etc.  It
should be noted that all theories for immiscible two-phase flow based
on refining the relative permeability approach will be unable to
handle this non-linearity.  Hence, it presents a huge challenge to the
porous media community.


{\sl Declaration} --- The authors declare no conflict of interest. 


{\sl Acknowledgement} --- The authors thank Dick Bedeaux, Carl Fredrik
Berg, Eirik G.\ Flekk{\o}y, Signe Kjelstrup, Knut J{\o}rgen
M{\aa}l{\o}y, Per Arne Slotte and Ole Tors{\ae}ter for interesting
discussions.  This work was partly supported by the Research Council of
Norway through its Centres of Excellence funding scheme, project
number 262644. 



\end{document}